# Human Creativity and Consciousness: Unintended Consequences of the Brain's Extraordinary Energy Efficiency?

by

T.N.Palmer
Department of Physics
University of Oxford

## Abstract

It is proposed that both human creativity and human consciousness are (unintended) consequences of the human brain's extraordinary energy efficiency. The topics of creativity and consciousness are treated separately, though have a common sub-structure. It is argued that creativity arises from a synergy between two cognitive modes of the human brain (which broadly coincide with Kahneman's Systems 1 and 2). In the first, available energy is spread across a relatively large network of neurons. As such, the amount of energy per active neuron is so small that the operation of such neurons is susceptible to thermal (ultimately quantum decoherent) noise. In the second, available energy is focussed on a small enough subset of neurons to guarantee a deterministic operation. An illustration of how this synergy can lead to creativity with implications for computing in silicon are discussed. Starting with a discussion of the concept of free will, the notion of consciousness is defined in terms of an awareness of what are perceived to be nearby counterfactual worlds in state space. It is argued that such awareness arises from an interplay between our memories on the one hand, and quantum physical mechanisms (where, unlike in classical physics, nearby counterfactual worlds play an indispensable dynamical role) in the ion channels of neural networks. As with the brain's susceptibility to noise, it is argued that in situations where quantum physics plays a role in the brain, it does so for reasons of energy efficiency. As an illustration of this definition of consciousness, a novel proposal is outlined as to why quantum entanglement appears so counter-intuitive.

1. ## Introduction

What does it mean to be human? Surely two of the defining characteristics are that of creativity and of consciousness (the latter being notoriously hard to define objectively). In this paper, a new perspective is speculatively proposed on the age-old problem of what it is physically about the human brain in particular which could account for these characteristics. In both cases, this proposal hinges around the notion of energy efficiency.

The author's interest in this subject arose by considering what must surely be one of the most profound paradoxes in computational neuroscience. For many years, attempts have been made to simulate parts of the brain on supercomputers. In the coming years, such simulations will start to use exascale high-performance computing. However, such computers will need 6 orders of magnitude more electrical power than the human brain itself needs (tens of megawatts rather than tens of watts). In Section 2, we review the arguments of Palmer and O'Shea (2015) that understanding this gross discrepancy in energy usage may provide a key clue to understanding the creative process, i.e. that creativity is an unintended consequence of the brain's evolution towards the extraordinarily energy efficient assemblage of miniaturised neurons that it is. In particular it is proposed that creativity arises from a very strong synergy between two cognitive modes of operation - one mode where limited available energy is focussed on a subset of neurons allowing these neurons to perform computations repeatably and reliably and where others are largely shut down, and the other mode where available energy is spread more uniformly around the neuronal network, making them susceptible to thermal noise, who existence in the brain is widely acknowledged (Faisal et al, 2008; Rolls and Deco, 2010). Implications for silicon computers are discussed in Section 3.

In Section 4, by analogy with thermal noise, we propose that quantum dynamics may play a role in the brain when it is energetically efficient to do so. As discussed, the power supply for action potentials in neural networks appears to be a case in point. We describe quantum physics as a theory where nearby counterfactual worlds in state space play a much more fundamental dynamical role than in classical physics and use this to give an explanation for our deeply held belief in free will.  In Section 5 we use this to provide a novel definition of consciousness. Here, we refer to consciousness as it relates to specific objects, i.e. in the sense of our being conscious of something. Being conscious of a specific object is then defined in terms of an ability to perceive that object as having an existence, independent of the rest of the field of view.  It is proposed that such consciousness arises from an interplay between our memories and the central dynamical role that nearby counterfactual states play in (e.g. the path integral representation of) quantum theory. As an illustration of this proposal, we discuss in Section 6, why we humans appear to find quantum entanglement so completely unintuitive.

## 2. Creativity

Consider the following very elementary example of creative thinking: Euclid's proof that there exists an unlimited number of prime numbers. Euclid starts by imagining the opposite: that the number N of primes is finite. This tactic is an essential step in many mathematical proofs (presumably even in Euclid's day) and has such broad application that one would hardly say it is an especially creative step for this particular problem. Indeed, it can readily be coded into a putative algorithm for finding mathematical proofs. The creative step comes in multiplying together the N primes and adding one. Having formed this number, it is immediate that it is not divisible by one of the N primes.

We will never know how Euclid (if indeed it was he) arrived at the idea of multiplying the primes together and adding one. Of course, in hindsight it seems so obvious that it is hard to imagine that one would do anything but this. However, conceivably Euclid may have started by adding the primes together, getting stuck, may have then multiplied them, getting stuck again, and may have given up, leaving his study to relax on the veranda. Then, in a moment of relaxation, the critical idea, "just add one" comes from nowhere. In an instant Euclid realises he had the proof he has been searching for.

The creative process has been described by the renowned mathematician J.E. Littlewood (following Helmholz and Poincaré):

"It is usual to distinguish four phases in creation: preparation, incubation, illumination and verification…..Illumination, which can happen in a fraction of a second, is the emergence of the creative idea into the consciousness. This almost always occurs when the mind is in a state of relaxation……." (Littlewood, 2004)

The same points have been emphasised by Andrew Wiles:

"In particular, when you reach a real impasse, when there's a real problem that you want to overcome, then the routine mathematical thinking is of no use to you. Leading up to that kind of new idea there has to be a long period of tremendous focus on the problem without any distraction. You have to really think about nothing but that problem - just concentrate on it. Then you stop. Afterwards there seems to be a kind of period of relaxation during which the subconscious appears to take over and it's during that time that some new insight comes." (Singh, 1997).

The notion that creative ideas come when relaxing, is commonplace. It seems to be an essential part of the creative process. But why?

The idea of changing from a mode of thinking where one focusses hard on a problem without distraction, to one where one simply relaxes, is suggestive of a switch in modes of cognition which Kahnemann (2012) refers to simply as "System 2" and "System 1" respectively. Kahnemann refers to System 2 as slow, effortful, logical, calculating; whilst System 1 is fast, automatic, frequent, emotional and stereotypic. Although it is simplistic to characterise cognition entirely in terms of such a dichotomy, it is conceptually convenient to do so here.

Here a physical reason for the difference between System 1 and System 2 is proposed, by returning to the astonishing fact that the brain performs exascale data processing with about 20W of power. The brain has achieved such extraordinary energy efficiency through the process of miniaturisation (Niven and Farris, 2012): the axon diameter of neurons is frequently as slender as 0.1 microns. Now whilst larger neurons (say of diameter 1 micron or more) are reliably deterministic and transmit information rapidly due to their low resistance to axial current flow, they are energy inefficient. In particular, larger neurons require more ionic current to trigger a nerve impulse. Following a bout of impulses, critical ionic concentrations across the neuronal membrane must be restored by energy-consuming ionic pumps. There would not be adequate energy to power sufficient neurons to allow the sorts of cognitive analysis humans undertake, if our neurons had 1-micron diameter. By contrast, smaller neurons are more efficient because their high input resistance allows relatively small trans-membrane ionic currents to generate the voltage needed to trigger a nerve impulse. The process of miniaturisation has allowed the human brain to contain around 80 billion neurons (by contrast chimpanzees have about 7 billion) with very limited energy resources.

There are potential disadvantages to such small neuronal diameters. Since signal transmission speed is smaller in slender neurons, reaction time to stimuli will be correspondingly slower. Hence, if fast reaction time is crucial for survival, then miniaturisation could be evolutionarily disadvantageous. However, with early hominids forming societal groups and learning to defend themselves collectively using primitive weapons, the need for ultra-fast reaction times may have started to become less important early in human history.

However, even though 20W may be sufficient to power a relatively large number of such slender neurons, it is still possible that it may not quite be sufficient to power a very large network of neurons so that they act repeatably and reliably in the presence of inevitable thermal noise (Faisal et al, 2008). For example, for very slender neurons, the ionic batteries positioned at key points along a neuron may each contain just a handful of ions. A consequence of limited energy is that the whole neuronal function can be susceptible to noise. Conventional neuroscience treats noise as an undesirable nuisance. For example, in their book "Principles of Neural Design", Stirling and Laughlin (2017) comment:

"Where noise is inevitable, it should be minimized before transmission, so most neural designs try to prevent noise or reduce it at early stages."

Whilst too much stochasticity would certainly be disadvantageous, a small degree of stochasticity could actually be advantageous, as Turing himself noted in his famous paper "The Imitation Game" (Turing, 1950). Deterministic heuristic algorithms for complex decision problems (e.g. the travelling salesman problem) can prove inefficient for particular problem instances, and in the worst case can lead to the algorithm "hanging" (Hoos and Stützle, 2005). Typically, there is a long "tail" in the distribution of problem instances where deterministic heuristics take an unacceptably long time to reach solution (Gomes et al, 1998). From an evolutionary perspective, such "Buridian donkey" behaviour would obviously be undesirable, if not fatal in the presence of predators. By eliminating this long tail, a stochastic heuristic algorithm can be more computationally overall than any corresponding deterministic algorithm (even though such stochastic algorithms may be slower than their deterministic counterparts for problem instances where the latter reaches solution rapidly). A particularly well known and successful stochastic algorithm is Simulated Annealing, used to find the global minimum of some objective function. The stochastic algorithm allows the search process to jump from the potential well surrounding a local minimum to the potential well surrounding the global minimum, in ways which deterministic algorithms would find difficult or impossible.

Here it is proposed that in System 2, available energy is focussed on a subset of neurons needed to perform a particular cognitive task as deterministically (and hence reliably and repeatably) as possible. This means that the limited energy available would be channelled to the specific parts of the brain needed to perform the computation, making this energy unavailable for performing other tasks. Kahneman notes that if you are out walking with a friend who suddenly asks you to multiply 23 by 17 (say), you may have to stop walking, close your eyes and essentially do nothing other than to focus on the task at hand. The neuronal processes needed to walk and talk and even process basic information coming from the sensory organs, have, *in extremis*, been switched off in System 2 mode. By contrast, in System 1 mode, one can happily walk, chew gum and simultaneously chat about the latest football results, since none of these activities requires unusual amounts of energy to be focussed on specific neuronal subsets.

This suggests that available energy per active neuron in System 1 mode is less that the available energy per active neuron in System 2 mode. We therefore postulate that because of this, it is specifically when in System 1 mode that neuronal action can be susceptible to thermal noise. Perhaps this is the reason why both Littlewood and Wiles comment that the creative moment comes when relaxing: it is in this state that a new idea can literally arise without prior reason. If we think of our present cognitive state in terms of a local minimum of some objective function whose global minimum defines the solution to the problem at hand, noise can take us out of the local minimum.

That is to say, perhaps the notion of "adding one" to the product of primes occurred to Euclid by means of a stochastic process. Perhaps the primitive stochastic idea was merely to "add some constant" and his System 2 immediately honed this down to adding one, rather than, say, two or ten. This serves to emphasise the notion that it is not simply the presence of noise that gives rise to creativity, it is the synergistic interplay between stochasticity and determinism associated with these two modes. This is consistent with Littlewood's comment that the process of illumination (a System 1 process) needs to be followed by the process of verification (a System 2 process). Put another way, it requires System 2 to determine whether the random jump from the local potential well will lead us to the global minimum. Using the language of particle physics, the mathematical physicist Michael Berry describes the product of the process of illumination as a clariton, but notes that, by virtue of the process of verification, the clariton is all too frequently annihilated by its anti-clariton partner (Ball, 2016). On top of which, the chances of stochastically jumping to the right answer becomes increasingly unlikely, the less prepared the brain is with the problem at hand – hence the vital roles of preparation and incubation.

One can feed students, or indeed program computers, with a "set of tricks" needed to prove mathematical theorems. "Just add one" could be one such trick. However, this trick is of no help in constructing a proof of perhaps the next simplest theorem in mathematics: that the square root of 2 is irrational. Indeed, one way of interpreting the Gödel-Turing theorem is that the set of tricks needed to prove mathematical theorems can never be taught. The fact that we humans are able to prove new mathematical theorems based on new mathematical tricks, could therefore be explained by the brain's susceptibility to noise in System 1. Indeed, this susceptibility provides a practical explanation of the Lucas/Penrose puzzle: How is it that we are able to see the truth of the Gödel theorem if our brains operate by algorithm? In the final analysis, for particularly slender human neurons, such noise may have its ultimate source in the supposed randomness of quantum decoherence. In this case, the randomness of the noise is as non-algorithmic as it can be. The possible dependence on quantum processes is further developed in Sections 4-6 below.

### 3. Lessons for Supercomputers

We may be able to learn in designing next-generation supercomputers by understanding how the brain has become so energy efficient. For example, part of the energy cost of operating a supercomputer is the cost of ensuring that the computations are bit reproducible – e.g. that computations are not affected by the internal effects of thermal noise or of external perturbations like cosmic rays. We can reduce energy costs by turning down the voltage across the transistors. However, in so doing they act less reliably (Palem, 2014). However, for a given unit of energy, one can ask what is more beneficial – a smaller number of precise computations, or a larger number of imprecise computations? The answer clearly depends on the application. In transferring sums of money from one bank account to another, it is clearly vital that precise bank account numbers are known. When adding or multiplying two floating-point real numbers, it is important to ensure the exponents are manipulated correctly. However, it is clearly less important for the trailing mantissa bits to be calculated precisely. For many computations in turbulent fluid mechanics (such is relevant for climate research for example), it is not vital that precise computations are made, for the simple reason that the numerical approximations to the underlying fluid equations are not themselves precise. To avoid systematically rounding errors, these trailing bits should be represented by noise, rather than systematically setting them to zero.

We have now reached the stage where it is no longer possible to shrink transistors any further and maintain complete determinism – energy dissipation would cause the circuitry to melt. As such, the current route to increased FLOP rates (floating point operations per second) is to combine more and more processors. The need to communicate across larger and larger networks of processors means that the energy costs are now dominated by the cost of transporting data from one processor to another (and to memory).

However, here we can perhaps learn from the brain. Energy need only be supplied to processors according to the required accuracy of the computation. That is to say, as in System 1 mode, we can turn the voltage down across the transistors where only imprecise estimates are needed. In this way, some computations will be susceptible to noise. As with the simulated annealing algorithm, sometimes this noise can be beneficial to the computation. In weather and climate model simulations, noise is certainly advantageous (Palmer 2019a). A consequence of this is that in moving data from one part of the computer to another, it is only be necessary to transport those bits that contain useful information which is distinguishable from noise. It is simply wasteful to transport those bits that are essentially indistinguishable from noise. This can reduce energy consumption considerably. Such an imprecise supercomputer (Palmer, 2015) can be compared with a typical energy-profligate scientific computation of today, where bit-reproducible arithmetic is computed using fixed precision (e.g. 64-bit floating point) real numbers. Currently there are no supercomputers with such imprecise capability, though the development of AI has meant that current computers are able to operate in some mixed precision mode, where 16- or even 8-bit variables can be efficiently processed.

This in turn raises the question of what it would need to make a (silicon) computer that is truly intelligent e.g. in the sense of performing interesting mathematical research. We have argued that intelligence arises from a complex interplay between stochasticity and determinism. It is not a matter of pre-programming the degree of stochasticity (vs determinism) in a fixed non-interactive manner. Rather this degree would itself have to be controlled by an extremely interactive operating system (e.g. which would perceive when a part of the code was effectively hanging and which part was making rapid progress in a purely deterministic manner). To make optimal use of available energy, stochasticity should be produced in hardware, rather than through pseudo-random number generators, and only data containing useful information should be transported within the computer.

## 4.  Quantum Physics, Counterfactuality, Free Will

It has been known for some years that quantum dynamics can play a key role in many biological systems (Al-Khalili, J. and J. McFadden, 2014). As an example, relevant to the current discussion, Summhammer et al (2018) consider the motion of $K^+$ ions in voltage-gated ion channels in the neuronal membrane wall. They note that it is difficult to explain the high rates of ion flow using classical physics: the potential barriers are too high according to the classical Nernst-Planck equation. A key observation is that the de Broglie wavelengths of such ions at typical brain temperatures are comparable with the scale of the periodic structure of Coulomb potentials in the nano-pore structure of the ion-channel selectivity filter.  Solving a nonlinear Schrödinger equation, Summhammer et al show that the ionic wavefunction can be sufficiently spatially spread that the front part of the wavefunction can effectively manipulate the confining potentials in such a way as to allow the remaining part of the wavefunction to propagate through. In this way, a mechanism for ion conduction has been found that would be impossible to achieve classically unless the ions had much larger kinetic energy (which would be impossible given the energy available to power neuronal dynamics). The characteristic timescale for the operation of this process has to be short, around 1ps, to explain the fast and directed permeation of ions through the potential barriers of the filter. In this way, the Summhammer et al mechanism not only builds on, but requires decoherence timescales of around 1ps, entirely consistent with the range of decoherence timescales associated with biological systems (a problem with many other hypotheses involving quantum physics in consciousness).

Consistent with the discussion above, it seems reasonable to hypothesise that the brain will use such a quantum process over a classical process when there is an energetic advantage to do so. Again, this becomes possible for the extreme miniaturisation of neurons in the human brain. However, what would be the consequences of this?

To answer this question, one needs to ask what is the key physical difference between quantum and classical physics. A clue arises from the fact that the essential quantum constant of nature, Planck's constant, has the dimension of position times momentum, i.e. the dimensions of the state space of a classical particle.  Classical theory can be thought of as arising in the limit when this dimensional constant is set equal to zero. This draws attention to the fact that a crucial difference between classical and quantum physics concerns the role of state space in the equations for dynamical evolution. In classical physics, a system's dynamical evolution between two given states is determined by the state-space trajectory along which the classical "action" (the integral of the Lagrangian along the trajectory) is minimised. In particular, trajectories (or so-called histories) which neighbour this path of least action play no direct role in determining the dynamical evolution of the system.

By contrast, in quantum theory dynamical evolution can be defined as a phase-weighted sum over trajectories in a region of state space (Feynman and Hibbs, 2010). Alternatively, in the De Broglie-Bohm representation of the Schrödinger equation, what makes the dynamics non-classical is the Bohmian quantum potential, a function, not in space-time, but on the configuration space of the system under investigation (Bohm and Hiley, 1993). Again, the presence of the quantum potential implies that one cannot isolate a single state-space trajectory when defining dynamical evolution. Rather, quantum dynamical evolution in the physical world is in some sense "aware" of the existence of alternative state-space trajectories in state space. Indeed the fact that quantum computers can outperform classical computers for certain problems can be viewed in terms of an ability of a quantum computer to exploit the parallelism implied by such neighbourhoods of state-space trajectories, in a manner impossible by a classical computer. If the physics which determines our cognition is "aware" of these neighbourhood trajectories, could our cognition itself be similarly aware?

From the perspective of some reference trajectory that we interpret as our physical world, such neighbouring state-space trajectories can be interpreted as counterfactual worlds: for example, worlds where some degrees of freedom are perturbed from the values that apply to the reference trajectory. Let us start by considering a "warm-up" for the problem of consciousness (discussed in the next section) - the deep-seated belief that many if not most of us have in the concept of free will. For many, this is the belief that "I could have done otherwise" (Kane, 2002). This definition supposes the meaningful existence of counterfactual worlds in which I did do otherwise. Perhaps in the counterfactual world I simply spent another fraction of a second looking to my right (keeping everything else in the world fixed). The consequences of this are easily deduced from our System 2 deductive reasoning: in this counterfactual world I would have seen the oncoming car, would not have pulled out of the turning, would not have collided with the car and would not have been in hospital for six months from where I write this paper (a fabricated example, fortunately). Whilst a world in which I didn't spend six months in hospital is distant from the world in which I did, a world where I spent an extra fraction of a second looking right resembles the actual world in almost all respects and therefore seems extremely close to the actual world. As such, it seems intuitively plausible to us to view this counterfactual world as physically reasonable (though in Section 6 we discuss why this sense of intuition may be misleading us).

Would I have such a strong belief in free will if the dynamics of our neural pathways was determined entirely by classical physics? For sure, it would be possible to refer back to past occasions (stored in my memory) where I did spend extra time looking left and right before turning out of some side road. However, these memories refer to different roads in different locations and certainly at different times. Does the memory of these past events explain the very visceral and deep-seated nature of the belief in the reality of the counterfactual world I looked right a fraction of a second longer for the particular road on the particular day when my accident occurred? I believe not. Of course, proving this unequivocally is impossible. Instead, using Littlewood's "preparation, incubation, illumination, verification" approach to creativity, the following assertion is made: that the notion of free will arises from a) a familiarity of different configurations of the world as stored in our memory (c.f. preparation and incubation), b) an awareness of nearby counterfactual worlds arising from the fact that (for reasons of energy efficiency) the dynamics of our neural pathways are partially influenced by quantum physical processes (c.f. illumination), c) an ability (using System 2) to project the consequences on such counterfactual worlds e.g. if I had looked right I wouldn't now be in hospital (c.f. verification). Here b) takes the place of the susceptibility to stochasticity, and indeed, if the origin of stochasticity is quantum decoherence, then there may be physical links between such processes in any case.

5. **Consciousness**

The literature on consciousness is so voluminous that no attempt is made to summarise it here. (See for example Schneider S. and M. Velmans, 2017). Instead, we attempt a novel definition of consciousness based on the ideas developed above. This section is necessarily very speculative in nature.

If I look at a bowl of fruit in the middle of the table, I can become conscious of it. What does this mean? Here we explore the idea that to be conscious of the bowl of fruit denotes an ability to treat the bowl of fruit as somehow distinct from and hence independent of the other objects in my field of view (the table, the walls of the room and so on). To treat the bowl of fruit as having an existence distinct from the other objects in my field of view, implies, in principle at least, an ability to perturb the degrees of freedom associated with the bowl of fruit (e.g. its position on the table) independently of the degrees of freedom of all the other objects in my field of view (i.e. keeping the latter fixed). That is to say, to be conscious of the bowl of fruit is to have some awareness of the existence of counterfactual worlds where the degrees of freedom of the bowl of fruit are perturbed relative to the degrees of freedom of all other objects in my field of view. If, moreover, I decide to focus on the contents of the bowl of fruit, I may become aware of the fact that it comprises, say, apples, oranges and pears. According to the definition above, to be conscious of a particular piece of fruit implies being aware, at least implicitly, of the existence of counterfactual worlds where the degrees of freedom of this particular piece of fruit, e.g. its position relative to the other pieces of fruit in the bowl, are perturbed.

From where does such awareness arise? Some would argue that it arises from the fact that in the past I have seen bowls of fruit in different positions on different tables, or of pieces of fruit arranged in different ways inside bowls of fruit. However, as in the discussion of free will, I do not believe that this explains the very visceral nature of consciousness. Again, a possible explanation is that our ability to perceive nearby counterfactual worlds arises from the fact that (for reasons of energy efficiency) quantum dynamics plays a central role in the operation of our neural networks, and just as quantum dynamics is aware of counterfactual worlds, so too our brains. It is then the interplay between this quantum-induced awareness of alternate worlds coupled with our memory of specific counterfactual worlds deemed close to the present one, that gives rise to the visceral nature of the experience we call consciousness.

A potential problem with such a definition is that it suggests that merely imagining something in one's mind's eye (e.g. a flying pig) could be enough to be conscious of that something. In general, that is not so, although under the influence of hallucinogenic drugs, people can apparently become viscerally conscious of illusory objects, suggesting that there may be a relatively fuzzy distinction between consciousness of real-world objects and imagined-world objects.

However, perhaps the fact that we are typically only conscious of real-world objects (rather than imagined-world objects) arises from the fact that amount of data transported from our sensory organs (e.g. along the visual cortex) is so much larger than for any other part of the brain's neural network. However, the simple volume of data may not be the relevant measure here. Instead perhaps the relevant diagnostic is the volume of data synchronised across nearby neurons (Singer, 1998). Here, a degree of synchronisation across the neurons of the visual cortex could conceivably be induced by the electromagnetic field associated with the action potentials of individual neurons (McFadden, 2002). Here, such synchronisation would be a semi-classical process, even though the ion flow in individual neurons is quantum mechanical.

In Section 2 we argued that creativity arises from a synergy between two modes of operation of the brain. It seems plausible to argue that a similar synergy arises in explaining the form of consciousness we call cognition. In her essay on consciousness, Magnusdottir (Magnusdottir, 2018) argues that the difference between "conscious experience" and "conscious cognition" is that the latter experience is supplemented with some predictive model. A predictive model need not be particularly sophisticated – it may simply be enough to enable a creature to anticipate the way a predator is likely to pounce. As such, by human standards, such a predictive model need not tax System 2 very greatly. However, it implies some computational effort in addition to a mere awareness of these counterfactual worlds. Hence, like creativity, it would seem that conscious cognition is also an interplay between a computational mode and a mode whose essential ingredient is either stochasticity in the case of creativity, or counterfactuality in the case of consciousness. It is interesting to note that both may have their origins in quantum physics. As Magnussdottir notes, simple perceptions and projections can be viewed as unconscious phenomena with and without (respectively) a predictive model. Here one could postulate that the synchronised data transport associated with such phenomena is not sufficiently great to trigger the type of quantum awareness considered here.

Some would argue that consciousness should be defined in terms of some kind of self-awareness. Indeed Magnusdottir (2018) herself asserts that consciousness requires a type of self-similar monitoring. However, the notion of self-awareness could be viewed as an extension of the more general notion of awareness of counterfactual alternatives described above. If I can perceive a bowl of fruit as having an existence independent of the rest of the world, so too I can perceive myself. However, as a subjective comment, I do not believe that for most of the time I am as actively conscious of myself as I am of the people and objects I interact with (unless I happen to look at myself in a mirror for example). This again is consistent with the notion that degree of consciousness of an object (which might include ourselves) is dependent on the (synchronous) transport of data which describes the object in question, along our neural pathways.

### 6. Why is Quantum Physics so Unintuitive?

As an application of the ideas above, consider the question of why quantum physics is so unintuitive. We focus on the phenomenon of entanglement on the basis that there is no more unintuitive idea in quantum physics than that of nonlocality. According to the Bell Theorem, any deterministic theory of quantum physics that satisfies the assumption of Statistical Independence (Hossenfelder and Palmer, 2019), must violate local causality. Hence, we seem to require that quantum physics must either be nonlocal (which the Bohmian formulation of quantum theory is) or is indeterministic (which the Copenhagen interpretation of quantum theory is). That is to say, we seem to be faced with the choice of spooky action at a distance, or of physics governed by randomness. Neither seems satisfactory.

The assumption of Statistical Independence ensures that when two or more sub-ensembles of quantum particles are being measured with different measurement settings (as occurs in a Bell experiment), each sub-ensemble is statistically similar to the others. If it were not possible to assume this then the basis for scientific investigation in general would be undermined.

However, the assumption of Statistical Independence does more than ensure that sub-ensembles are statistically independent, it implies a strong form of counterfactual definiteness (Hossenfelder and Palmer, 2019). Consider two measurements performed on two sub-ensembles of particles: counterfactual definiteness assumes that in principle one could have performed the second measurement on the first sub-ensemble (even though in practice one did not).

As we have speculated, our intuition about counterfactual words arises from an interplay of quantum physical processes in the brain, together with our past experiences. In terms of this, any counterfactual world which is sufficiently close to the real world, is a plausible world. However, this notion of closeness assumes a metric on state space, and it is natural to assume the familiar Euclidean metric. Such an assumption is the product of the deepest of all our intuitions. As babies, we try to put nearby brightly coloured objects into our mouths (in case they are a source of food or indeed drink). To do this we must bring the objects close to our mouths. We learn by trial and error what it means for an object to be close to our mouths. We learn by empiricism about the Euclidean metric of space. This is our go-to metric.

However, there is no reason for the Euclidean metric to be the correct metric of distance in state space rather than physical space. In particular, there is a deterministic theory of quantum physics based on fractal geometry, where the relevant metric on state space is p-adic rather than Euclidean (Palmer, 2019b). Relative to this metric, putative states which lie in the gaps in the fractal geometry are necessarily distant from points on the fractal set, even when a fractal gap appears very slender and hence insignificant from a Euclidean perspective. In a theory where states of physical reality necessarily lie on this fractal set, then counterfactual worlds which do not lie on the fractal set are not physically realistic.

By ruling out such counterfactuals we can violate Statistical Independence without violating the statistical independence of sub-ensembles of particles occurring in reality. This suggests that the reason we find quantum physics so difficult to understand is because our intuition, that <u>all</u> counterfactual worlds which sufficiently well resemble the real world are plausible worlds, is false.

From this perspective, the "weirdness" of quantum physics arises from a cognitive inability to discriminate between those counterfactual states of the world which are realistic and plausible and those which are not. To make better sense of contemporary physics it is better to avoid using counterfactuals at all, e.g. by saying that I am free if there are no constraints preventing me from doing what I want to do (Kane, 2002). Certainly, such real-world definitions are vital if we are to explain the violation of the Bell inequality in a causal theory where experimenters are free agents (Palmer, 2019b; Hossenfelder and Palmer, 2019).

## 7. Conclusions

It is postulated that, through its evolution over many millions of years, the miniaturisation of the neuronal pathways in the brain has resulted in an exceptionally energy efficient organ. The disadvantages of such miniaturisation (e.g. relatively slow reaction times in the presence of predators) has been offset by two advantages: an ability to be creative and an ability to be aware of the world around us. It is argued that those advantages arise from two specific manifestations of energy efficiency: the role of stochastic noise and the role of quantum parallelism respectively.

As such, it seems plausible to speculate that it will be impossible to replicate human intelligence in strictly deterministic algorithmic machines, as suggested by Penrose (1994). However, by trying to solve the energy efficiency problem in high performance computing, we may find that we start to produce machines which have intelligence characteristics more akin to those of humans.